\documentstyle[epsfig]{ioplppt}

\def\epm#1#2{\hbox{${+#1}\atop {-#2}$}}

\def\gsim{\mathrel{\rlap{\lower4pt\hbox{\hskip1pt$\sim$}}
    \raise1pt\hbox{$>$}}}         

\def\etal{{\it et al.}}

\def\frac#1#2{{{#1}\over {#2}}}
\def\half{\hbox{${1\over 2}$}}\def\third{\hbox{${1\over 3}$}}

\def\smallfrac#1#2{\hbox{${{#1}\over {#2}}$}}

\def\GeV{{\rm GeV}}
\def\MS{\hbox{$\overline{\rm MS}$}}

\catcode`@=11 
\def\slash#1{\mathord{\mathpalette\c@ncel#1}}
 \def\c@ncel#1#2{\ooalign{$\hfil#1\mkern1mu/\hfil$\crcr$#1#2$}}
\def\lsim{\mathrel{\mathpalette\@versim<}}
\def\gsim{\mathrel{\mathpalette\@versim>}}
 \def\@versim#1#2{\lower0.2ex\vbox{\baselineskip\z@skip\lineskip\z@skip
       \lineskiplimit\z@\ialign{$\m@th#1\hfil##$\crcr#2\crcr\sim\crcr}}}
\catcode`@=12 

\def\PR{{\it Phys.~Rev.~}}
\def\PRL{{\it Phys.~Rev.~Lett.~}}
\def\NP{{\it Nucl.~Phys.~}}

\def\PL{{\it Phys.~Lett.~}}

\def\ZP{{\it Zeit.~Phys.~}}

\def\APP{{\it Acta.~Phys.~Pol.~}}
\def\EPJ{{\it Eur.~Phys.~Jour.~}}
\def\vol#1{{\bf #1}}\def\vyp#1#2#3{\vol{#1}, #3 (#2)}

\begin{document}

\title{Spin Physics}

\author{R.D.~Ball\dag\S\addtocounter{footnote}{-1}
\footnotetext{\S Royal Society University Research Fellow} 
and H.A.M$^c$L.~Tallini\ddag}

\address{\dag Department of Physics and Astronomy, University of Edinburgh,\\
Edinburgh EH9 3JZ, Scotland}
\address{\ddag Department of Physics, University of Liverpool,\\
Liverpool L69 3BX, England}

\begin{abstract}
We review the current situation in polarized scattering experiments. 
We describe the theoretical interpretation of inclusive deep 
inelastic processes, 
the current experimental situation and perturbative analyses which extract 
structure functions and parton distribution functions. We also discuss
various issues such as positivity constraints, small $x$ and the possibility
of polarized colliding beam experiments at HERA, semi-inclusive processes,
the separation of flavors and the measurement of the gluon polarization, 
and the possibility of using polarisation experiments to put 
constraints on new physics.
\end{abstract}


\section{Introduction}

In a polarized DIS experiment, both target and beam spins are 
longitudinally polarized.
Analogously to writing the spin averaged DIS cross section in terms of two
unpolarized structure functions, the spin dependent differential cross
section
can be written in terms of two spin structure functions, 
$g_1(x,Q^2)$ and $g_2(x,Q^2)$:
\begin{equation}
\frac{d^2 \sigma^{+}}{dxdy}-\frac{d^2 \sigma^{-}}{dxdy}=
\frac{8\pi \alpha ^2ME}{Q^4}
\left\{ \left( 2y -y^2 - \frac{Mxy^2}{E} \right) 2x{g_1} 
  -\frac{4M}{E}x^2y{g_2} \right\},\label{xsec}
\end{equation}
where $d^2 \sigma^{\pm}$ are the differential cross sections for target
and beam spins respectively parallel and antiparallel. 
The longitudinal inclusive asymmetry is defined as
\begin{equation}
{A_1} = \frac{d^2 \sigma^{+}_{\gamma N} - d^2 \sigma ^{-}_{\gamma N}}
{d^2 \sigma^{+}_{\gamma N} + d^2 \sigma ^{-}_{\gamma N}}  
={\frac{g_1-\gamma ^2 g_2}{F_1}}
=(1+\gamma^2)\frac{g_1}{F_1}-\gamma A_2.\label{asym}
\end{equation}
As the kinematic factor ${\gamma={2Mx}/{\sqrt{Q^2}}}$ is usually
small, and $A_2$ is bounded by $\sqrt{R}$, the 
longitudinally polarized structure function can be approximated by 
$g_1 \approx A_1F_1$.

Just as the unpolarized structure function $F_1$ has a simple
interpretation in the parton model as the sum over the contributions of the 
two helicity states of quarks and antiquarks,
\begin{equation}
F_1(x)= \half \sum_{i} e^2_i (q_i^{+}+\bar{q}_i^{+}+ q_i^{-}+\bar{q}^{-}_i) =
\half\sum_{i} e^2 q_i(x),\label{upart} \\
\end{equation}
where $q_i$ are the unpolarized quark plus antiquark
distribution functions of flavour $i$ in
the nucleon, so the polarized structure function $g_1$ is interpreted as the
difference,
\begin{equation}
g_1(x)=\half \sum_{i} e^2_i (q_i^{+}+\bar{q}_i^{+} 
- q_i^{-}-\bar{q}^{-}_i ) =\half
\sum_{i} e^2 \Delta q_i(x),\label{polpart} \\
\end{equation}
where $\Delta q_i(x)$ are the 
polarized quark plus antiquark distributions.

In perturbative QCD, this identification of parton distributions with 
structure functions is modified by perturbative corrections:
\begin{eqnarray}
F_1(x) &= \half\langle e^2\rangle [C^{\rm NS}\otimes q_{\rm NS}
     +C^{\rm S}\otimes q_{\rm S}
     +C^{\rm g}\otimes g]\label{factu} + O(1/Q^2)\\
g_1(x) &= \half\langle e^2\rangle [\Delta C^{\rm NS}\otimes \Delta q_{\rm NS}
     +\Delta C^{\rm S}\otimes \Delta q_{\rm S}
     +\Delta C^{\rm g}\otimes \Delta g] + O(1/Q^2),\label{factp} 
\end{eqnarray}
where $q_{\rm NS}$, $\Delta q_{\rm NS}$ are flavor nonsinglet combinations 
of unpolarized and polarized quark plus antiquark densities, 
$q_{\rm S}$, $\Delta q_{\rm S}$
the corresponding flavor singlet quark densities, and $g$ and $\Delta g$ the
unpolarized and polarized gluon densities. These are convoluted in the 
usual way with the quark coefficient functions $C^{\rm NS}$,$C^{\rm S}$,
$\Delta C^{\rm NS}$,$\Delta C^{\rm S}=1+O(\alpha_s)$ and gluon 
coefficient functions 
$C^{\rm g}(Q^2)$,$\Delta C^{\rm g}(Q^2)=O(\alpha_s)$, derived from hard
cross-sections expanded perturbatively (at NLO) in $\alpha_s(Q^2)$. The parton 
densities are evaluated at scale $Q^2$: their
evolution with $t=\ln Q^2$ is then given by the Altarelli-Parisi equations
\begin{eqnarray}
\frac{d}{dt}q_{\rm NS}
= P^{\rm NS}_{qq} \otimes q_{\rm NS},\nonumber\\
\frac{d}{dt}\ \left (\matrix{q_{\rm S} \cr g}\right)
=\left (\matrix{P_{qq}^S & P_{qg}^S \cr P_{gq}^S
& P_{gg}^S }\right) \otimes \left (\matrix{q_{\rm S} \cr g}\right),
\label{APu}\\
\frac{d}{dt}\Delta q_{\rm NS}
= \Delta P^{\rm NS}_{qq}
\otimes \Delta q_{\rm NS},\nonumber\\
\frac{d}{dt}\ \left (\matrix{\Delta q^{\rm S} \cr \Delta g}\right)
=\left (\matrix{\Delta P_{qq}^S & \Delta P_{qg}^S \cr \Delta P_{gq}^S
& \Delta P_{gg}^S }\right) \otimes 
\left (\matrix{\Delta q^{\rm S} \cr \Delta g}\right),
\label{APp}
\end{eqnarray}
where the various unpolarized and polarized splitting functions are
expanded perturbatively: at NLO $P$,$\Delta P = O(\alpha_s)+O(\alpha_s^2)$.
Note that unpolarized and polarized, and nonsinglet and singlet distributions,
evolve independently, while the singlet quark distributions mix with the
corresponding gluon distributions.

All of the coefficient functions and splitting functions have now been 
computed at NLO, the polarized calculations being completed only 
recently \cite{NLO}. These calculations were performed using 
\MS\ subtraction, with a particular definition of $\gamma^5$, and with a 
further finite subtraction to ensure conservation of first moments (see 
\cite{NLO} for details). Of course one can however use them to do 
calculations in any scheme one chooses by making a NLO redefinition of 
parton densities. The choice of scheme is of particular relevance for 
the first moments of the polarized densities.

The first moment of $g_1$ for proton and neutron targets can be written 
(assuming exact isospin symmetry and ignoring charm for simplicity) as
\begin{displaymath}
  \Gamma^{p(n)}_1 =
\smallfrac{1}{12}C^{\rm NS}(\pm a_3+\third a_8)
  +\smallfrac{1}{9}(C^{\rm S}\Delta\Sigma +C^{\rm g}\Delta g) ,\label{fmgen}
\end{displaymath}
where $\Delta\Sigma$, $a_3$ and $a_8$ are the first moments of  
singlet and non-singlet quark plus antiquark distributions,
\begin{equation}
a_3=\Delta u - \Delta d,\qquad a_8=\Delta u + \Delta d - 2 \Delta s,\qquad
\Delta\Sigma=\Delta u + \Delta d + \Delta s,\label{heldef} 
\end{equation}
$\Delta g$ is the first moment of the polarized gluon distribution, 
and $C_{\rm S}$, $C_{\rm NS}$  and $C^{\rm g}$ are the first 
moments of the corresponding coefficient functions. In a general scheme
the first moments of the quark distributions $\Delta q_i$ will depend on 
renormalization scale. If we wish to identify them with the quark 
helicities, it is necessary to adopt schemes in which they are renormalization 
group invariant \cite{AR,SV}: this is possible only if each 
of $a_3$, $a_8$ and $\Delta\Sigma$ is scale independent. 
In all such schemes it can be shown 
\cite{AR} that the first moment of the gluon coefficient function 
$C^g=-n_f\frac{\alpha_s}{2\pi}C^{\rm S}$, whence
\begin{equation}
  \Gamma^{p(n)}_1(Q^2) =
\smallfrac{1}{12}C^{\rm NS}(\alpha_s(Q^2))(\pm a_3+\third a_8)
  +\smallfrac{1}{9}C^{\rm S}(\alpha_s(Q^2))a_0(Q^2),\label{fmhel}
\end{equation}
where the axial singlet charge
\begin{equation}
  a_0(Q^2) =
\Delta\Sigma-n_f\frac{\alpha_s(Q^2)}{2\pi}\Delta g(Q^2).\label{axsing}
\end{equation} 
Because of the Adler-Bardeen nonrenormalization theorem reasonable 
schemes exist in which these equations hold to all orders in 
perturbation theory \cite{BFRb}: such schemes have become known 
as `AB schemes'.\footnote{
An example of such a scheme was constructed in ref.\cite{BFRb}: others
are provided by using jets to define the polarized gluon distribution  
ref.\cite{AR}, provided sufficient care is taken to absorb all soft 
contributions into the parton densities \cite{jets}.}  
Naively one might have thought that the second term in eqn.\ref{axsing} 
is small asymptotically, but this is not the case because $\Delta g(Q^2)$
grows linearly with $\ln Q^2$ \cite{AR}. It follows that for first moments
of polarized distributions the partonic identification eqn.\ref{polpart} 
is unjustified; polarized gluons have an effectively pointlike coupling 
to polarized virtual photons through the triangle anomaly.

The Bjorken sum rule~\cite{bjsumrule} follows directly from eqn.\ref{fmhel} 
and the observation that in the schemes in which $a_3$ is scale independent it
may be identified with the forward nucleon matrix element of the 
(partially conserved) axial current, and thus with the ratio axial coupling 
$g_A$ measured in neutron $\beta$-decay:
\begin{equation}
\Gamma_1^p-\Gamma_1^n 
=\smallfrac{1}{6}\left(1-\smallfrac{\alpha_s}{\pi}-\ldots\right)g_A.
\label{Bj}
\end{equation}
It is thus a fundamental prediction of perturbative QCD.

Adopting the more dangerous assumption that the
strange quarks in the nucleon are unpolarized, and moreover that the
gluon polarization is sufficiently small that the distinction eqn.\ref{axsing} 
between $\Delta\Sigma$ and $a_0(\infty)$ may be ignored, 
leads to the `Ellis-Jaffe sum rule'~\cite{ejsumrule},
\begin{equation}
\Gamma_1^{p(n)}=
\pm\smallfrac{1}{12}\left[
1-\smallfrac{\alpha_s}{\pi}
+\smallfrac{5}{3}\smallfrac{3F-D}{F+D}
\left(1-\smallfrac{7\alpha_s}{15\pi}\right)+O(\alpha_s^2)
\right]g_A,
\label{EJ}
\end{equation}
where the ratio $F/D$ is deduced from hyperon decays\cite{Rat}. It was the
observation by EMC of the violation of this sum rule, from which it 
was deduced that the axial singlet charge $a_0$ was much smaller than 
the octet charge $a_8$, that rekindled interest in polarized 
DIS in 1987~\cite{EMC}. Two distinct interpretations of this result 
were suggested: either $\Delta s$ is unexpectedly large (and negative)
but $\Delta g$ is small so that $a_0\simeq\Delta\Sigma\ll a_8$, 
or $\Delta s$ is indeed small, but $\Delta g$ is large (and positive) 
so that $a_0\ll\Delta\Sigma\simeq a_8$. The basic question still 
remains as to how the spin of the nucleon is divided up among 
its constituents:
\begin{equation}
  \half =  \half {{(\Delta u + \Delta d + \Delta s + \ldots)}} +\Delta g+ L_z,
\end{equation}
where the scale dependence of $L_z$, the contribution from orbital 
angular momentum \cite{orb}, precisely cancels that of $\Delta g$.
Since the EMC results were published, several second generation
experiments were proposed to make more accurate measurements in polarized
DIS, and most of these now have results. 

\section{Experimental programmes and Inclusive Asymmetries}

\begin{table}

\caption{Summary of polarized DIS data taken to date.}
\label{tab:expts}
\lineup
\begin{indented}
\item[]\begin{tabular}{||c|c|c|c|c||} 
\br
Expt & Beam E & Target & x-Bjorken \\
\mr
E80(1975)~\cite{E80} & 16 GeV e$^-$ & H-butanol & 0.1--0.5 \\
E130(1980)~\cite{E130} & 16-23 GeV e$^-$ & H-butanol & 0.2--0.7 \\
E142(1992)~\cite{E142} & 19-26 GeV e$^-$ & $^3$He & 0.03--0.6 \\
E143(1993)~\cite{E143} & 10-29 GeV e$^-$ & NH$_3$/ND$_2$ & 0.03--0.8 \\
E154(1995)~\cite{E154} & 49 GeV e$^-$ & $^3$He & 0.014--0.7 \\
E155(1997)~\cite{E155} & 49 GeV e$^-$ & NH$_3$/LiD & 0.014--0.85 \\
\mr
EMC(1985)~\cite{EMC} & 100-200 GeV $\mu^+$ & NH$_3$ & 0.005--0.75 \\ 
SMC(1992)~\cite{SMC92} & 100 GeV $\mu^+$ & D-butanol &
0.003--0.7 \\
SMC(1993)~\cite{SMC93} & 190 GeV $\mu^+$ & H-butanol &
0.003--0.7 \\
SMC(1994)~\cite{SMC945} & 190 GeV $\mu^+$ & D-butanol &
0.003--0.7 \\
SMC(1995)~\cite{SMC945} & 190 GeV $\mu^+$ & D-butanol &
0.003--0.7 \\
SMC(1996)~\cite{SMC96} & 190 GeV $\mu^+$ & NH$_3$ & 0.003--0.7
\\
\mr
HERMES(1995)~\cite{HERMES95} & 28 GeV e$^+$ & $^3$He & 0.02--0.7 \\
HERMES(1996) & 28 GeV $^+$ & H & 0.02--0.7 \\
HERMES(1997)~\cite{HERMES97} & 28 GeV $^+$ & H & 0.02--0.7 \\
\br

\end{tabular}
\end{indented}
\end{table}

\begin{figure}[hbt!]
\vskip -0.7truecm
\vbox{
\hbox{\hskip 5.0truecm
\hfil\epsfxsize=7.1truecm\epsfbox{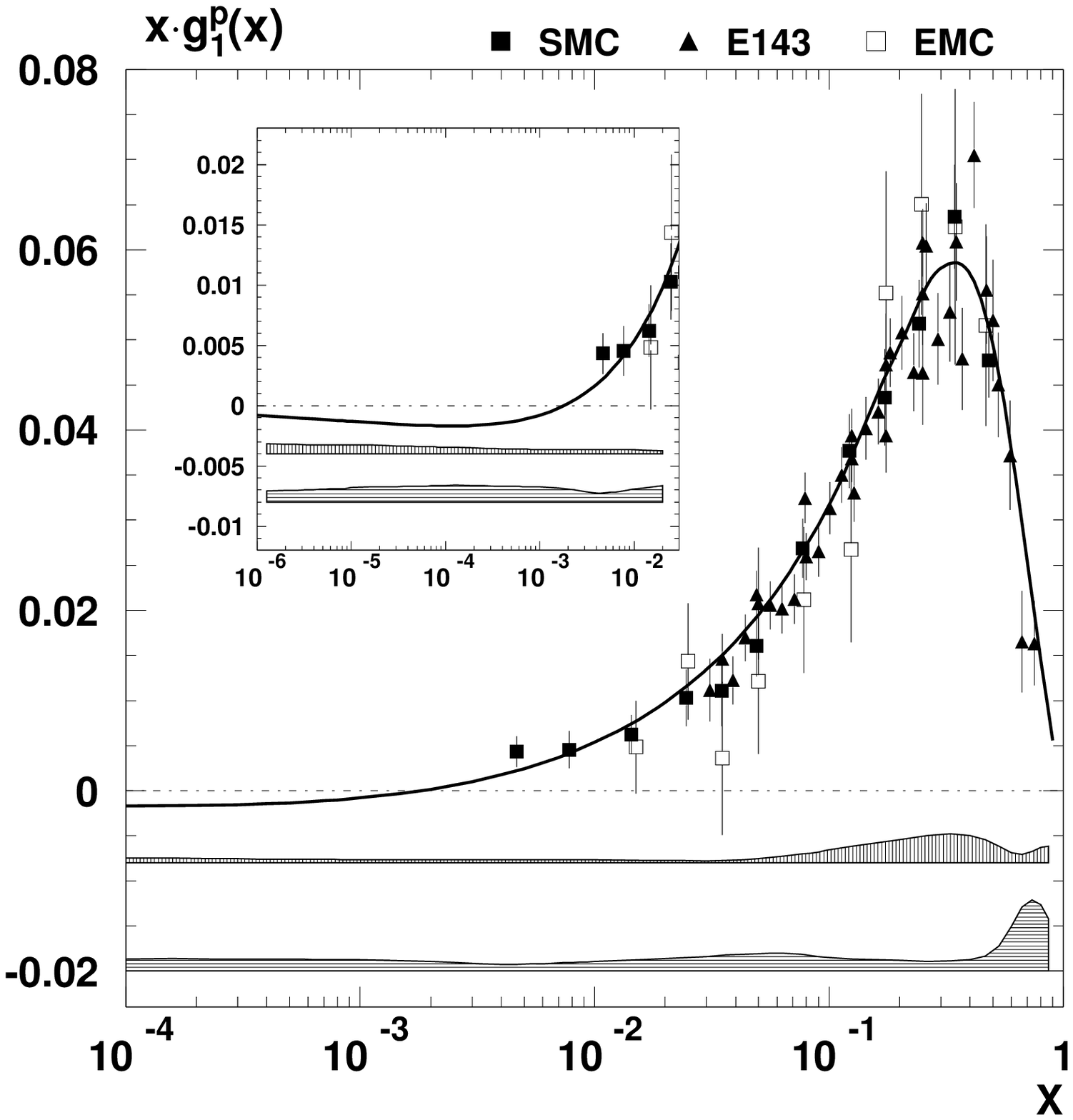}\hfil}
\vskip 0.0truecm
\hbox{\hskip 5.0truecm
\hfil\epsfxsize=7.1truecm\epsfbox{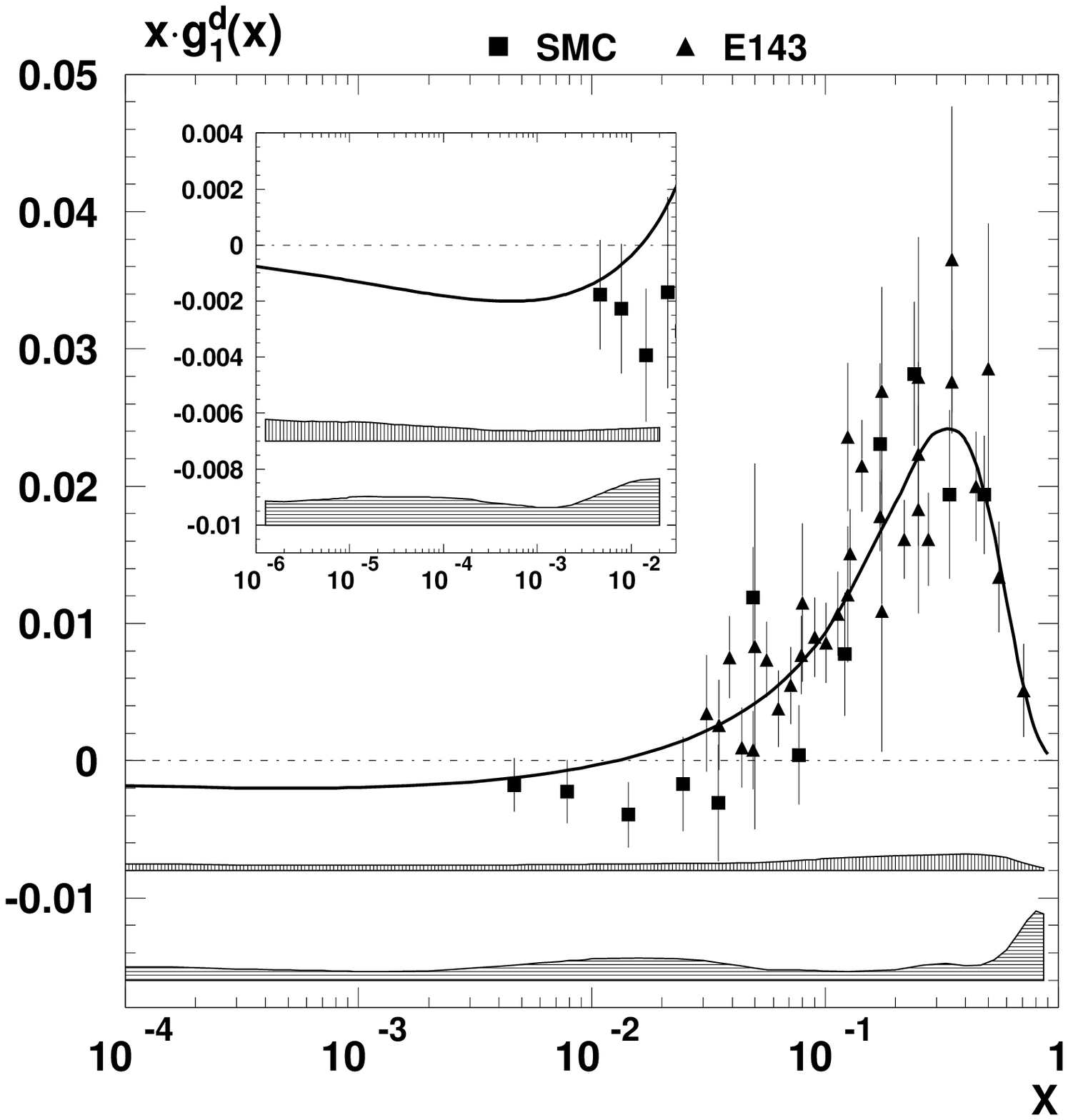}\hfil}
\vskip 0.0truecm
\hbox{\hskip 5.0truecm
\hfil\epsfxsize=7.1truecm\epsfbox{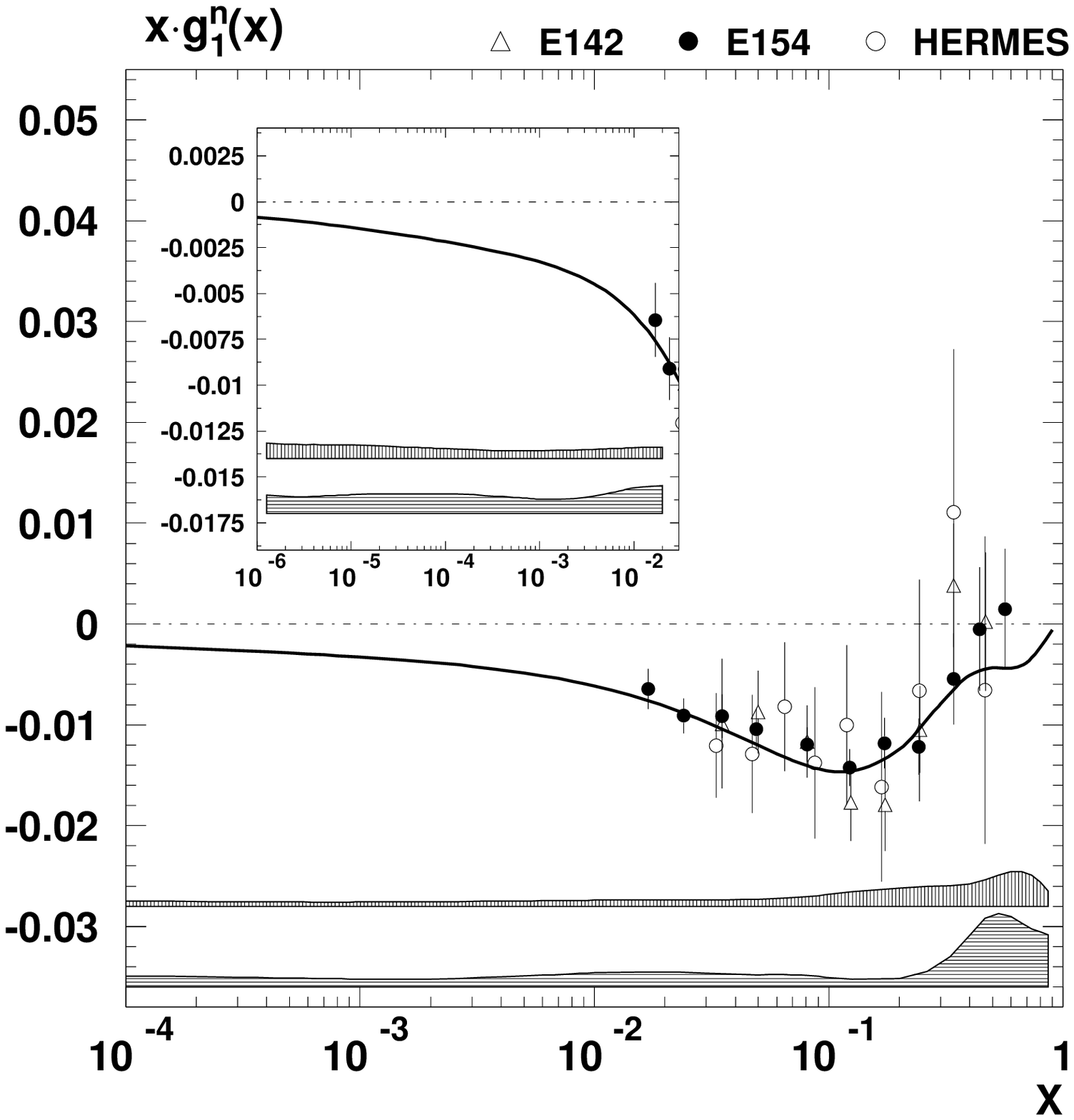}\hfil}}
\vskip 0.3cm
\caption[]{\label{g1fit}
Data on $g_1^p$, $g_1^d$ and $g_1^n$. The error bars are statistical: the 
upper error band gives the systematic errors. Also shown is the global 
SMC fit \cite{SMCfit} at $Q^2=5\GeV^2$, with a theoretical error band 
(lower). The insets show the small $x$ region.}
\end{figure}

\begin{figure}[hbt!] 
\setlength{\unitlength}{1cm}
\begin{center}
 \begin{picture}(10.,7.0)
  \put(-2.5,1.){\epsfig{file=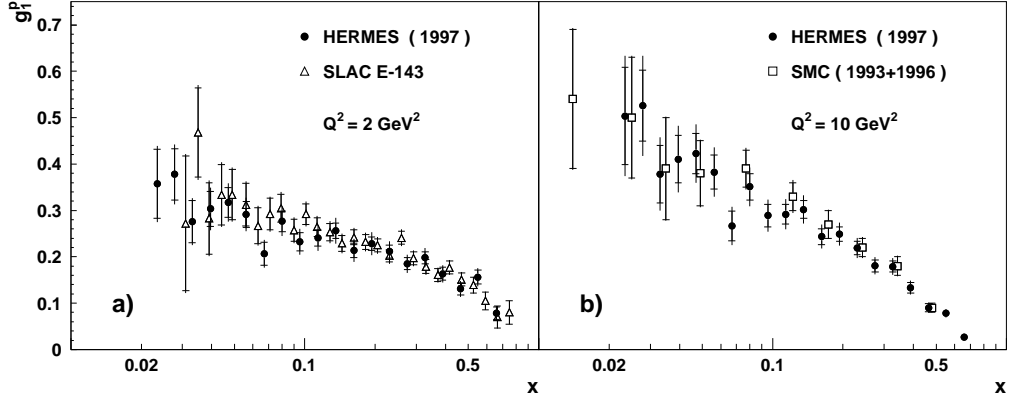,width=0.7\textheight}}
 \end{picture}
\end{center}
\vskip-2.0cm 
\caption{\label{fig:hermes_g1}
The HERMES $g_1^p$ results evolved to 2~$GeV^2$ and 10~$GeV^2$
for comparison with E143 and SMC data.}
\end{figure}

\begin{figure}[hbt!]
\setlength{\unitlength}{1cm}
\begin{center}
 \begin{picture}(10.,7.0)

\put(-2.0,7.55){\epsfig{figure=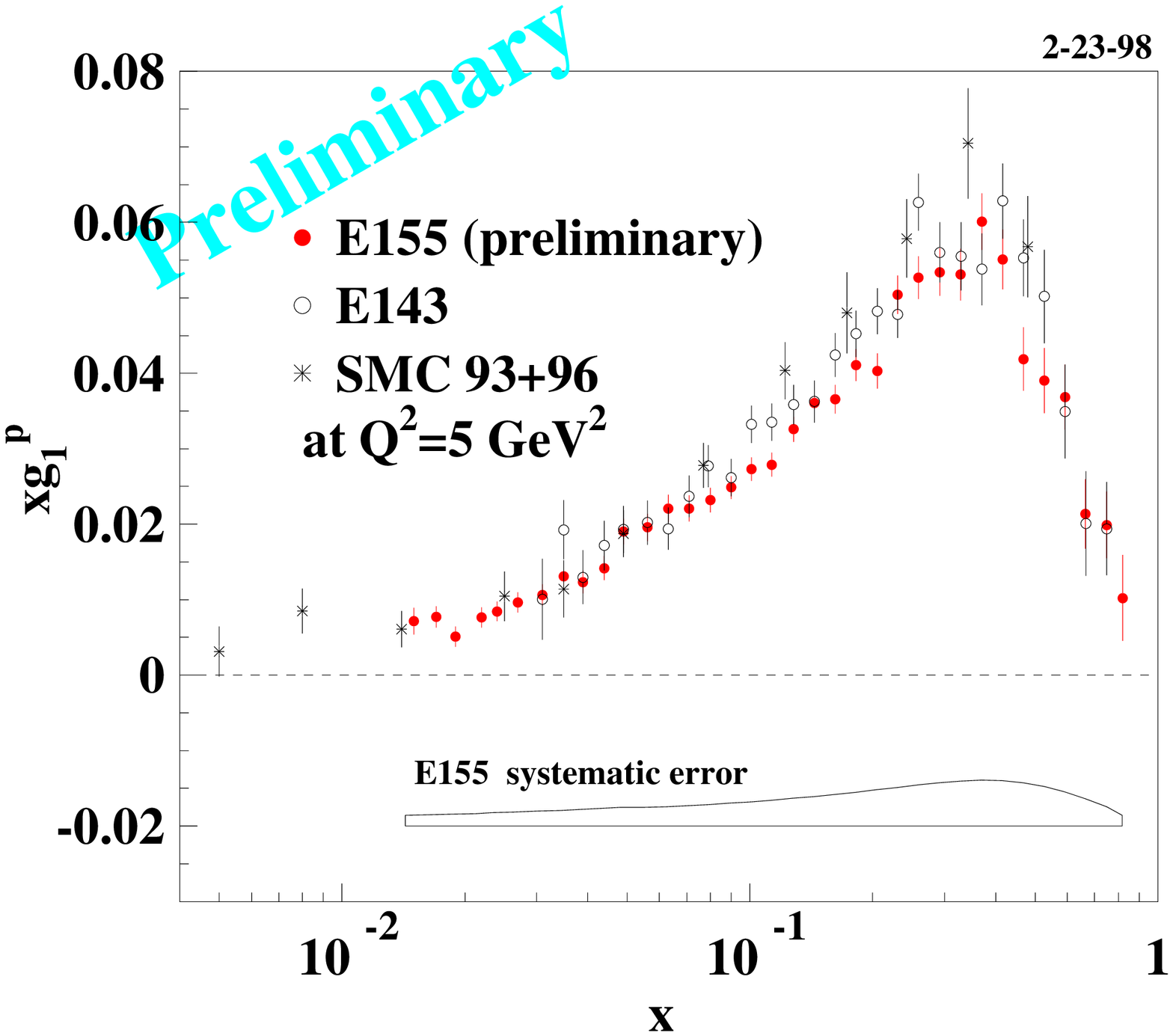,width=0.45\textwidth,angle=270}}

\put(5.5,7.){\epsfig{figure=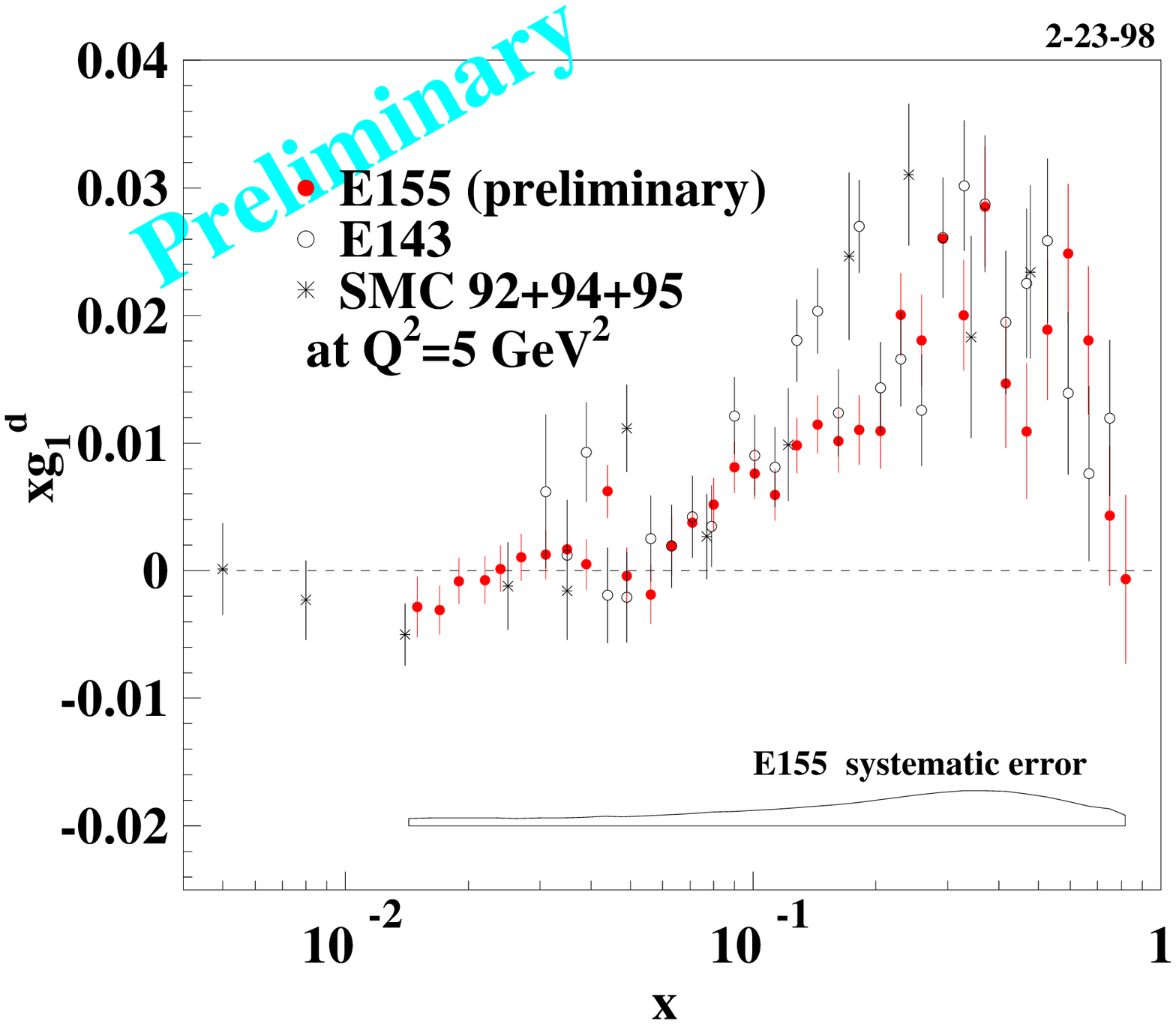,width=0.420\textwidth,angle=270}}
 \end{picture}
\end{center}
\vskip -2.0truecm
\caption{\label{fig:e155_g1} 
E155 preliminary results on $g_1$ for (a) the proton and (b)
deuterium targets.}
\end{figure}

A summary of the data taken by all polarized DIS experiments to date is
presented in table~\ref{tab:expts}. The experimental programmes are 
complementary. SMC uses the highest energy, though least intense, beam
and hence has the largest range in $x$ and $Q^2$. This large kinematic 
reach is vital for the measurement of first moments, where 
the uncertainty of the polarized structure function extrapolation at low
$x$ is large.
The SLAC experiments do not get to $x$ values as low as SMC, 
but the high intensity linear accelerator they use allows them to produce
results with excellent precision.

Currently HERMES is the only experiment taking data.
It uses a novel experimental technique of gaseous polarised targets 
in an electron (positron) storage ring.
It also has excellent particle identification capability which enables it 
to make a broad range
of semi-inclusive as well as inclusive measurements.
Gas targets are advantageous because they have a large  proportion of 
polarisable nucleons which can polarised to a high degree.
Though pure neutron targets do not exist, a model of a
neutron can be used instead, for example $^3$He or deuterium.

Most of the older datasets are displayed in figure~\ref{g1fit}.
The HERMES, SMC and E143 results on $g_1^p$ are compared in
figure~\ref{fig:hermes_g1}, and are seen to be consistent.
Currently the most accurate inclusive data comes from the E155
experiment, whose preliminary results for proton and deuterium targets 
are shown in figure~\ref{fig:e155_g1}. These results represent a fraction
of the data, with data at high $Q^2$ still being analysed.

\section{The Extraction of Polarized Parton Distributions}

The analysis of polarized structure function data using NLO perturbative QCD
proceeds along just the same lines as that of unpolarized data. An
ansatz for the polarized distributions at a starting scale $Q_0^2$ 
is chosen (in this case $\Delta q_{\rm NS}$, $\Delta q_{\rm S}$ and 
$\Delta g$), this is evolved up to the data using the evolution equations 
eqn.\ref{APp}, $g_1^N$ is calculated using eqn.\ref{factp}, and compared to 
the data.
The initial parameterization is then tuned in order to find the best fit.
Such calculations have been performed by various groups over the last few years
\cite{BFRb,SMCfit,BFRa,polpart,fm,Leader}. 

Here we will describe the results of the recent SMC analysis \cite{SMCfit}. 
This uses the code developed in \cite{BFRb,BFRa}, supplemented by a 
proper treatment of experimental systematic errors. All except the most 
recent HERMES and E155 data in figures~\ref{fig:hermes_g1} 
and~\ref{fig:e155_g1} was included in a global analysis.
The fitted curves are displayed in figure~\ref{g1fit}. The areas under 
the curves give the first moments of $g_1$ and from these one can extract 
(using eqn\ref{fmhel}) 
\begin{eqnarray}
&a_3 = 1.20 \epm{0.08}{0.07}{\rm (stat.)}
\pm 0.12 {\rm (syst.)}\epm{0.10}{0.04}{\rm (th.)}\\
&a_0(\infty) = 0.24 \pm 0.07 {\rm (stat.)}\pm 0.19{\rm (syst.)}.
\label{axe}
\end{eqnarray} 
The first of these results confirms the Bjorken sum rule eqn.\ref{Bj} at 
around the $10\%$ level (since from $\beta$-decay $g_A = 1.259$); the 
second confirms the violation of the Ellis-Jaffe sum rule (since 
from hyperon-decay $g_8 = 0.58$\cite{Rat}). Indeed the result for the 
axial singlet charge is again compatible with zero, just as it was 
for the original EMC analysis \cite{EMC}. It is also possible to determine
$\alpha_s$ from scaling violations (though not, as yet, from 
the Bjorken sum rule): the SMC global fit gives
\begin{equation}
\alpha_s(M_Z^2) = 0.121 \pm 0.002{\rm (stat.)}\pm 0.006 {\rm (syst. \& th.)},
\end{equation}
consistent with an earlier analysis\cite{fm} and the world average of 
$0.118\pm 0.003$.

\begin{figure}[hbt!]
\vbox{
\hbox{\hskip 2.5truecm
\hfil\epsfxsize=12.0truecm\epsfbox{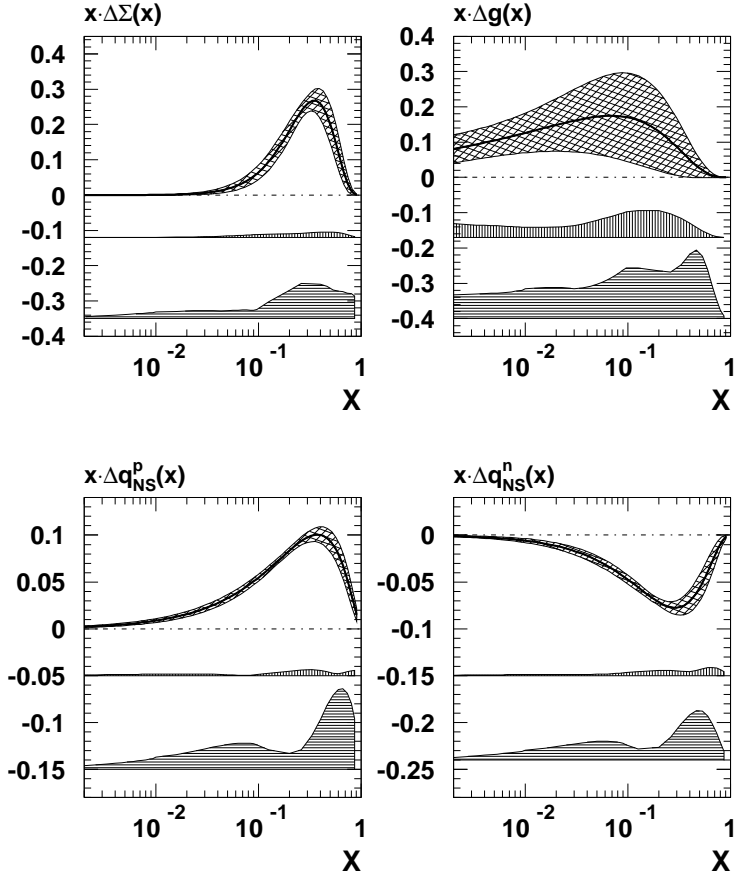}
\hfil}}
\vskip 0.5cm
\caption[]{\label{SMCpdfs}
The parton distribution functions $\Delta\Sigma\equiv\Delta q_{\rm S}$,
$\Delta g$ and $\Delta q_{\rm NS}$ (for both proton and neutron) 
obtained from the SMC fits \cite{SMCfit}. All the distributions are at 
$Q_0^2=1\GeV^2$ and in the AB-scheme of ref.\cite{BFRb}. 
The three error bands are statistical, systematic and theoretical 
respectively.
}
\end{figure}

The polarized parton distributions obtained from these global fits are shown 
in figure~\ref{SMCpdfs}. Besides a central distribution, the authors made 
a full error analysis (something which, incidentally, would also be 
invaluable in global fits of unpolarized distributions~\cite{errors}).
The error bands in the figures must be treated with caution, since 
correlations between errors at small and large $x$ are not shown. The 
theoretical error includes an estimate of NNLO corrections from variation 
of the renormalization and factorization scales by a factor of $\sqrt 2$,
and an estimate of the dependence on the choice of parameterization. 
Target mass corrections~\cite{TMC} and higher twist corrections are not 
included however. For all the distributions the theoretical errors are 
at least as big as the experimental errors, and at large $x$ are 
generally much bigger.

From fig.~\ref{SMCpdfs} it is clear that both $\Delta q_{\rm S}(x)$ and 
$\Delta g(x)$ are significantly positive and of comparable size (though 
of course the errors on the former are much smaller). This confirms the 
discovery of a positive $\Delta g$ first noted in ref.\cite{BFRb}.
On the other hand the results obtained for the first moments,
\begin{eqnarray}
\Delta\Sigma &= 0.38 \pm 0.03 {\rm (stat.)}
\epm {0.03}{0.02}{\rm (syst.)}\epm{0.03}{0.05}{\rm (th.)}\\
\Delta g &= 1.0 \epm{1.2}{0.3} {\rm (stat.)}\epm{0.4}{0.2}{\rm (syst.)}
\epm{1.4}{0.5}{\rm (syst.)}
\label{sfm}
\end{eqnarray} 
suggest that there might also be a sizeable strange quark 
component (since $\Delta\Sigma$ seems significantly lower than 
$a_8=0.58$).\footnote{Some caution is required here, however, since 
the fitted $\Delta q_{\rm S}(x)$ is strongly valencelike at \
small $x$, so the small $x$ uncertainty may be underestimated.}

Comparison of $\Delta q_{\rm NS}^{\rm p}$ with   
$-\Delta q_{\rm NS}^{\rm n}$ shows a small discrepancy which can be
attributed to the octet contribution (see also ref.\cite{Leader}). 
Note however that this discrepancy is considerably smaller than 
the estimated theoretical error, so it is still difficult \cite{fm} to 
use it to confirm the value of $F/D$.

\section{Positivity Bounds}

Since the differential cross-sections eqn.\ref{xsec} are positive, 
the asymmetry eqn.\ref{asym} must be bounded by unity because of 
the triangle inequality: if $d\sigma^{\pm}\geq 0$,
\begin{equation}
0\leq |d\sigma^{+}-d\sigma^{-}|\leq d\sigma^{+}+d\sigma^{-},
\label{triangle}
\end{equation}
whence $|A_1|\leq 1$. In the parton model the parton distributions 
are simply proportional to the differential cross-sections 
(eqns.\ref{upart},\ref{polpart}), and thus the inequality eqn.\ref{triangle}
translates into the `positivity bound'
\begin{equation}
0\leq |f^{+}-f^{-}|\leq f^{+}+f^{-},
\label{parttriangle}
\end{equation}
or $|\Delta f|\leq f$ for each charged parton (so $f$ can 
correspond to any of the $q_i$ or $\bar{q}_i$).\footnote
{For usual DIS with virtual photon exchange, the positivity bound 
is strictly speaking only derivable for the 
particular parton combinations eqns.\ref{upart},\ref{polpart}. However if 
we allow chiral exchange currents which couple separately to each flavour, 
the more restrictive (but equally natural) bounds 
eqn.\ref{parttriangle} are required.} 

Consider a simplified situation in which there is only one type of parton.
Then in NLO perturbative QCD we may write schematically
\begin{eqnarray}
d\sigma^{+} &= (1+\alpha_s c^{+})\otimes f^{+} + 
\alpha_s c^{-}\otimes f^{-},\nonumber\\
d\sigma^{-} &= (1+\alpha_s c^{+})\otimes f^{-} + \alpha_s c^{-}\otimes f^{+},
\label{nlopmxsec}
\end{eqnarray}
where $c^{+}$ results from a NLO hard scattering process which preserves 
helicity, $c^{-}$ from one that flips helicity. The positivity of the 
cross-sections now implies that 
\begin{equation}
f^{+}\geq -\alpha_s c^{-}\otimes f^{-} +O(\alpha_s^2),\qquad
f^{-}\geq -\alpha_s c^{-}\otimes f^{+} +O(\alpha_s^2).
\end{equation} 
It follows that the partonic positivity conditions $f^{\pm}\geq 0$
or either relaxed or tightened depending on whether the spin flip 
partonic cross-section $c^{-}$ is positive or negative. In terms of 
unpolarized and polarized distributions 
\begin{equation}
|\Delta f|\leq 1+\alpha_s(c-\Delta c)\otimes f +O(\alpha_s^2),
\label{nlotriangle}
\end{equation}  
where $c=c^{+}+c^{-}$, $\Delta c=c^{+}-c^{-}$. At large scales
the NLO bound eqn.\ref{nlotriangle} goes over to the LO bound
eqn.\ref{parttriangle}, but conversely at low scales they can be 
significantly different.

There is nothing mysterious here: partons are not physical objects, 
but rather scheme 
dependent theoretical constructions. In a `parton scheme' $c^{\pm}=0$ and 
$f^{\pm}\geq 0$ to any order. But in other schemes positivity bounds 
on parton distributions will generally be nontrivial beyond LO.\footnote{Note 
however that these 
considerations are independent of the definition of first moments: 
positivity constraints are of no consequence for first moments since while 
$\Delta f$ remains finite, $f$ diverges.} It follows that it only makes sense
to impose partonic positivity bounds eqn.\ref{parttriangle} in 
partonic schemes.
The fits which incorporate such bounds \cite{polpart} thus run the risk 
of being over-constrained, particularly when they also evolve from scales 
so low that the use of perturbation theory is unreasonable \cite{fm}.

A full numerical analysis of NLO positivity bounds requires a 
physical cross-section which is proportional to $\Delta g$ at leading order,
for example Higgs production in polarized gluon-nucleon scattering, 
evaluated at NLO \cite{pos}. The results confirm that the bounds are 
only useful at values of $x\gsim 0.3$, where the measured 
asymmetry tends to one. The distributions obtained from the SMC fit 
\cite{SMCfit} lie well inside the bounds, despite the fact that they were not
imposed in the analysis. The usefulness of positivity bounds as a constraint 
on the polarized gluon at very large $x$ is hampered by the large 
uncertainties in the unpolarized gluon distribution \cite{gluun}, 
and indeed by the lack of a complete error analysis in global fits to 
unpolarized data.

\section{Polarized Partons at Small $x$}

In order to compute first moments of polarized structure functions or 
parton distributions, it is necessary to extrapolate the measured 
distribution from the measured region down to $x=0$. At one time this was 
done using Regge theory \cite{Heimann}: this predicts that $g_1$ will be 
flat or valencelike at small $x$, and consequently that the contribution to
first moments will be small and under control. However Regge theory is only 
expected to apply at low $Q^2$. Perturbative corrections at larger values of 
$Q^2$ generate instabilities due to contributions from logarithms of $1/x$ 
to the evolution eqns \ref{APu},\ref{APp}. For unpolarized distributions
the most important small $x$ singularity is that in the triple gluon 
vertex \cite{DeRuj}: this drives a dynamical rise in $F_2$ at 
small $x$ which grows as $Q^2$ increases, which is clearly seen in the data
from HERA \cite{DAS,BdR}. In the polarized case all of the splitting 
functions contain singularities \cite{prise}, which mix in such a way 
that $g_1$ is inevitably driven negative at small $x$ with 
increasing $Q^2$ \cite{BFRa,Erice}. It follows that when first moments are 
measured in deep inelastic scattering, the small $x$ contribution is actually
rather large and negative, as may be seen from the fitted curves in 
figure~\ref{g1fit}. The theoretical uncertainty in the perturbative 
extrapolation may be conservatively estimated by varying scales: it 
is inevitably rather large. Thus current estimates of first moments 
tend to be rather lower than some of the earlier ones, with a larger 
(and indeed now dominant) uncertainty coming from the small $x$ extrapolation
(see table~\ref{tab:sx}). 

\begin{table}[b!]
\vskip-0.5cm
\caption{Contributions to first moments of $g_1^N(x,Q_0^2)$ at 
$Q_0^2=5\GeV ^2$ \cite{SMCfit}. Regge extrapolation of the neutron data
makes little sense, as the data themselves show no indication of 
Regge behaviour.}
\label{tab:sx}
\begin{center}
\begin{tabular} {||l|c|c|c|}
\br  & $0.003<x<0.8$ & $0.0 < x <0.003$  & $0.0 < x <0.003$  \\
&  (meas)& (fit) & (Regge) 
\\
\mr
{}Proton & $0.130\pm 0.003\pm 0.005\pm 0.004$ & 
-0.012\epm{0.014}{0.025} & $0.002\pm 0.002$  \\
{}Deuteron & $0.036\pm 0.004\pm 0.003\pm 0.002$ 
& -0.015\epm{0.010}{0.023} & $0.0\pm 0.005$(?) \\
{}Neutron & $-0.054\pm 0.007\pm 0.005\pm 0.004$
& -0.020\epm{0.010}{0.026} & ? \\
\br
\end{tabular}
\end{center}
\end{table}

\begin{figure}[t!]
\vskip-1.5cm
\vbox{\hbox{\hskip 0.5truecm
\hfil
\epsfig{file=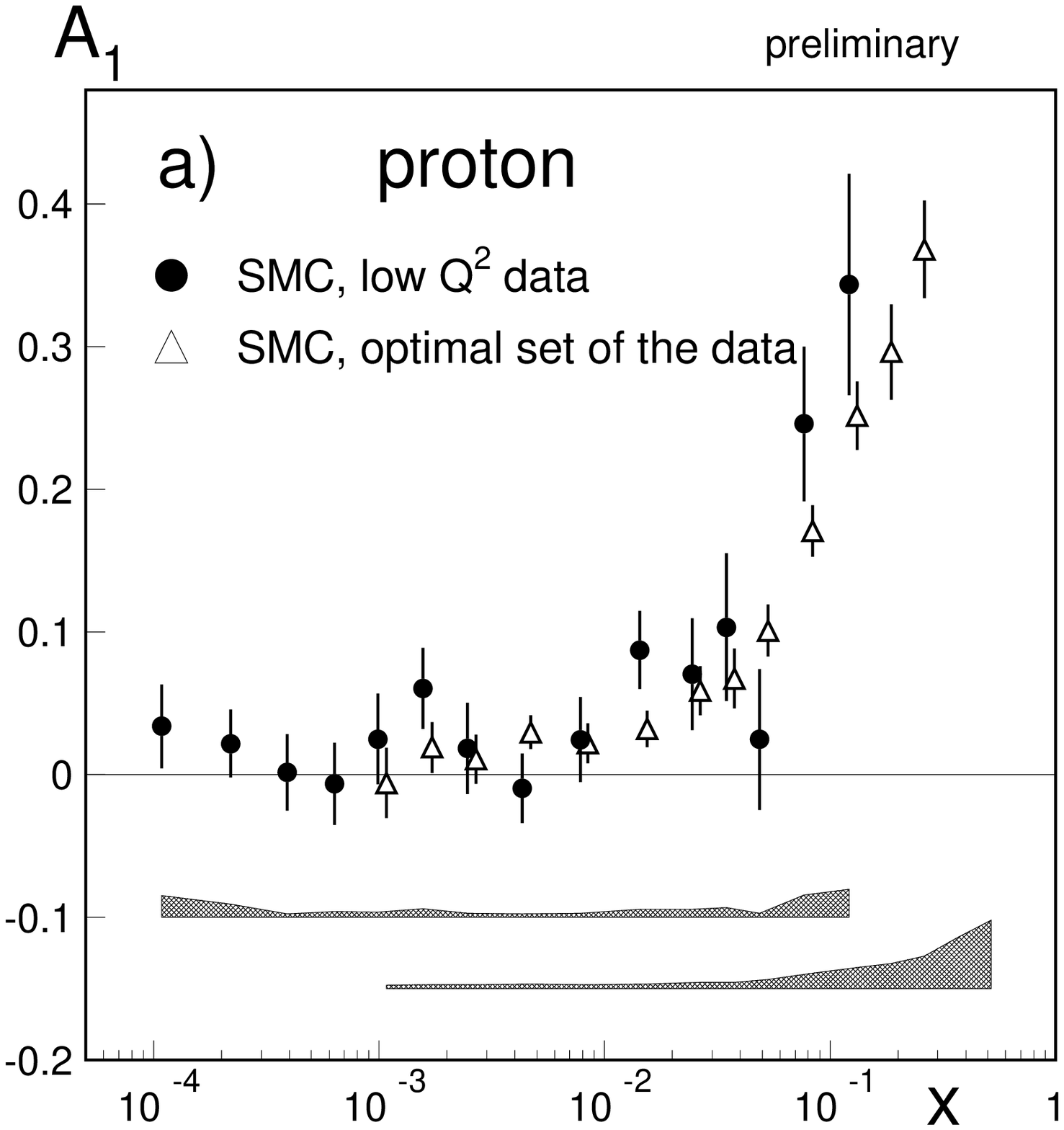,width=7cm,
bbllx=0pt,bblly=100pt,bburx=600pt,bbury=650pt}
\hskip 0.0truecm
\epsfig{file=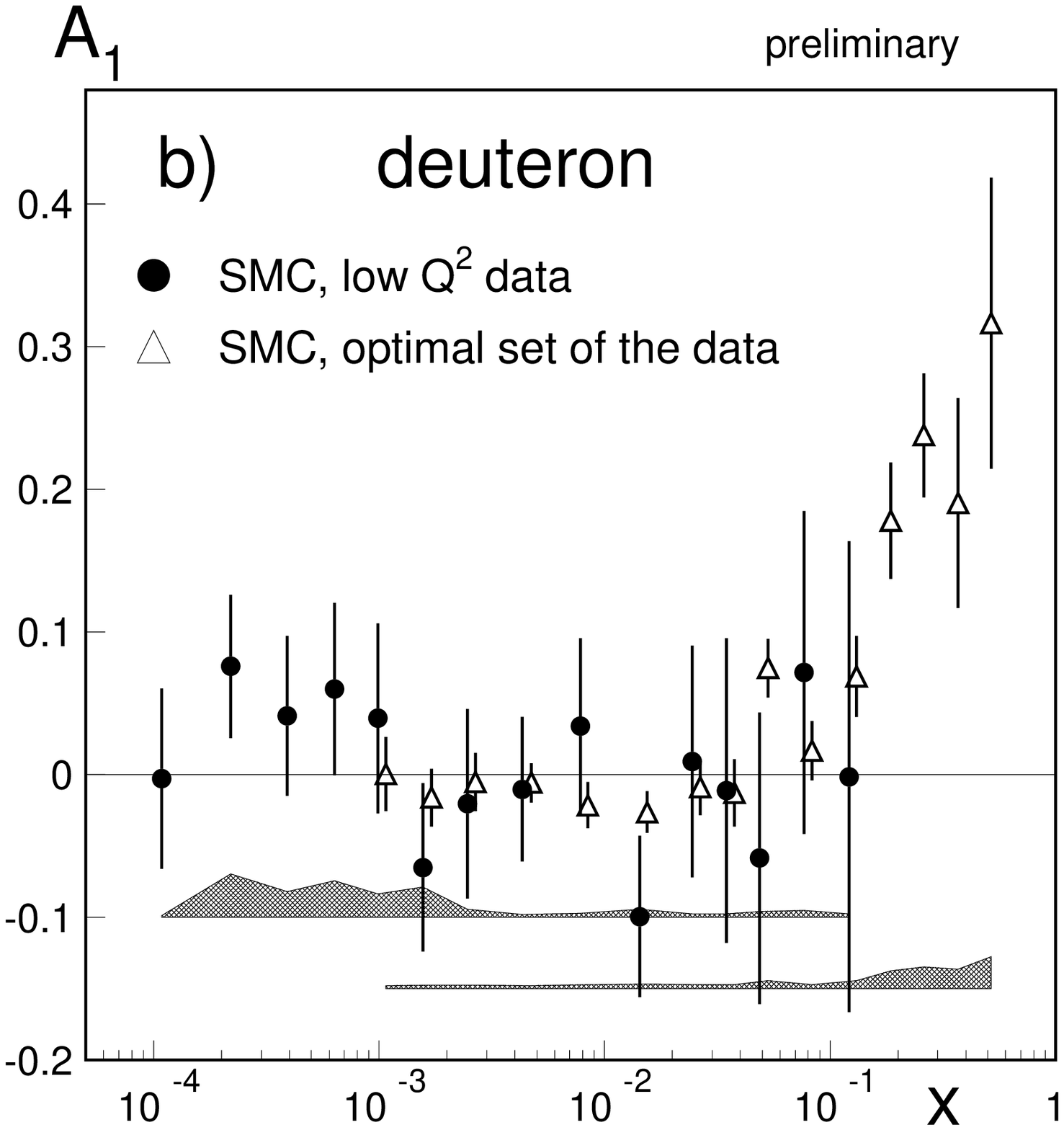,width=7cm,
bbllx=0pt,bblly=100pt,bburx=600pt,bbury=650pt}
\hfil}}
\vskip 0.8cm
\caption[]{\label{SMCsx}
SMC data at small $x$ and low $Q^2$ \cite{SMCsxlq}.
}
\end{figure}

To reduce this uncertainty it would be very useful to have more data.
The current experimental situation on $A_1^p$ at small $x$ is summarised in 
figure~\ref{SMCsx}. The four points with $x\lsim 10^{-3}$ have $Q^2$ between
$0.01\GeV^2$ and $0.2\GeV^2$, which should be well inside the Regge region.
The results are consistent with a flat or valencelike distribution, 
very dramatic rises at small $x$ being ruled out, though it is 
difficult to say much more. Indeed $g_1$ is very difficult to measure at 
small $x$ since $A_1\sim xg_1/F_2$, so even if both $g_1$ and $F_2$ 
were approximately flat (as expected from Regge theory),
the asymmetry would still be falling as $x$, suggesting that at 
$x\sim 10^{-4}$ it could be smaller than one per mille. 

\begin{figure}[b!]
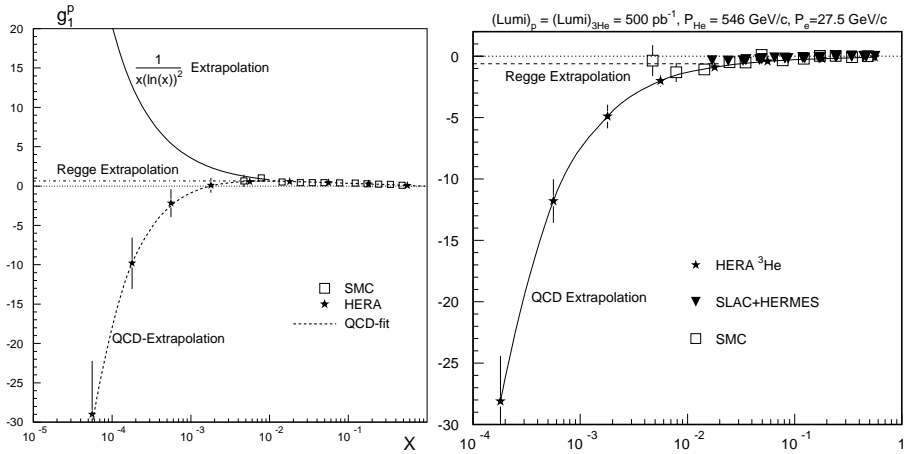

\vbox{\hbox{\hskip 0.5truecm
\hfil\epsfig{file=g1pHera.epi,width=56mm}
\hskip 0.0truecm
\epsfig{file=g1dHera.epi,width=62mm}
\hfil}}
\vskip 0.5cm
\caption[]{\label{g1Hera}
Projected measurements of $g_1^p$ and $g_1^d$ in colliding beam experiments
at HERA, assuming a total luminosity of $500{\rm pb}^{-1}$ \cite{HERAS}.
}
\end{figure}

Various studies have been carried out to investigate the exciting 
possibility of
measuring $g_1^p$ at small $x$ at HERA in collider mode \cite{HERAS}.
In order to do this the proton beam would have to be polarized: this may 
be technically feasible, indeed polarized proton beams are being 
planned for RHIC \cite{RHIC}, but is certainly a 
difficult and costly exercise. 
Making the optimistic assumption that a luminosity of $500{\rm pb}^{-1}$
is obtainable (current luminosities with an unpolarized proton beam are 
an order of magnitude smaller than this) an idea of the possible accuracy
for a measurement of $g_1^p$ is shown in figure~\ref{g1Hera}. Such a 
measurement would substantially reduce the error in the first moment 
of $g_1^p$, and could confirm the general trend of the QCD extrapolation.
To improve the precision on the Bjorken sum, necessary if it is to be used 
to determine $\alpha_s$ at NNLO, and improve the precision on the 
first moment of $\Delta g$, it would be necessary to measure in addition 
$g_1^n$. This could be done with a beam of polarized He$^3$.

Besides the small $x$ logarithms included in the NLO perturbative 
calculations described above there are further logarithms of higher orders:
in polarized parton evolution these are `double' logarithms of the form
$\alpha_s^n\ln^{2n-1}x$. Summing them up gives a very steep perturbative rise
\cite{KL,BER} (as a power of $1/x$, which in the singlet case threatens to 
spoil the convergence of first moments). It has been suggested that one
could search for such behaviour at a polarized HERA. In practice this would be 
very difficult: the sign of the powerlike rise is the same as that of 
the more conventional perturbative behaviour, which it simply reinforces.  
Double scaling plots \cite{DAS} would have very few points, with large errors.

In any case our experience in the unpolarized case suggests that summing up
small $x$ logarithms in this way is not so easy. Here the double 
logarithms cancel, leaving only single logarithms of the 
form $\alpha_s^n\ln^{n}x$, and it is then possible to extend 
conventional collinear factorization to
all orders even in the small $x$ limit \cite{CH}. 
However recent calculations of the 
next-to-leading-logarithms \cite{FL} (i.e. those of the form 
$\alpha_s^n\ln^{n-1}x$) show that the expansion in small $x$ logarithms
is simply not very useful for values of $\alpha_s$ as large as $0.1$ 
\cite{NLLx}. Various studies of the NLO BFKL equation reveal instabilities
leading to negative cross-sections \cite{NLOBFKL}. Given this confusion,
it would take considerable courage to put much faith in double logarithm 
summations in the polarized case.

\section{Flavour Decomposition and Semi-inclusive Asymmetries}

In semi-inclusive measurements a final state
hadron may be used to tag the flavour of the struck quark. Using this extra
information, it is possible to measure the separate contributions to the
nucleon's spin from each quark flavour.

\begin{figure}[hbt!]
  \setlength{\unitlength}{1cm}
  \begin{center}
    \begin{picture}(8.,6.)
      \put(0.,-.5){\epsfig{figure=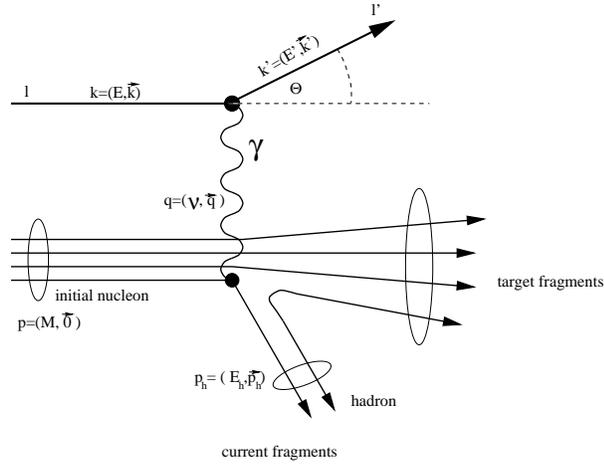,width=8cm}}
    \end{picture}
  \end{center}
  \caption{\label{fig:semi} 
Schematic diagram of a semi-inclusive DIS event.}
\end{figure}

\begin{figure}[hbt!] 
  \setlength{\unitlength}{1cm}
  \begin{center}
    \begin{picture}(8.,10.)
      \put(-1.0,-0.5){\epsfig{figure=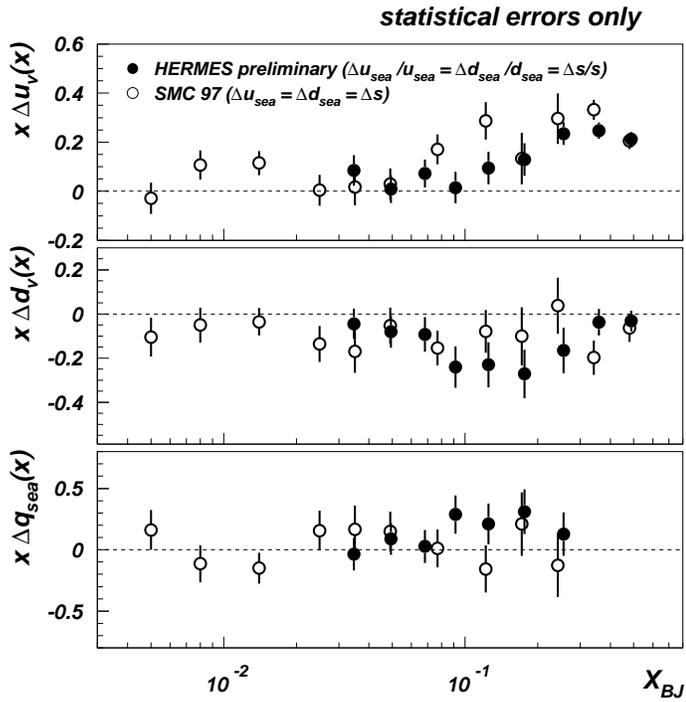,width=9cm}}
    \end{picture}
  \end{center}
  \caption{\label{fig:deltaq} 
Polarized quark distributions from SMC complete data
 set~\cite{SMC_semi} and HERMES preliminary 
results based on the first two years of data taking~\cite{HERMES_semi}.}
\end{figure}

A schematic diagram for the process is shown in figure~\ref{fig:semi}. 
Provided that factorisation holds, the cross section for the production 
of a particular hadron $h$ is
\begin{equation}\label{eq:semixc}
\frac{d \sigma^h(z)}{dz}=\frac{\sum_i e_i^2
q_i(x)D_i^h(z)}{\sum_ie^2q_i(x)}\sigma^T(x),
\end{equation}
where $\sigma^T$ is the total inclusive cross section and $D_i^h(z)$ are
the fragmentation functions, which represent the probability of a struck quark
of flavour $i$ forming a hadron of type $h$. 
The spin asymmetry for a hadron $h$ can then be written as
\begin{equation}
  A_N^h \equiv \frac{{\Delta}{\sigma}^h_N}{{\sigma}^h_N} = 
  \frac{{\sum} e^2_i{{\Delta}q_{i}(x)} D^h_i(z)}
  {{\sum} e^2_iq_{i}(x) D^h_i(z)}.
\end{equation}
The polarized quark distributions can then be extracted from this
equation. 

Currently only two experiments, SMC and HERMES, are capable of making
semi-inclusive measurements and have performed analyses to extract the
polarized quark distributions. 
To maximise the information available, fits are made to positive and
negative hadron asymmetries and to inclusive asymmetries for both proton and
neutron targets. To reduce the number of unknowns an SU(3) flavour 
symmetric sea is assumed. The unpolarized parton distributions are 
taken from existing parameterisations. 
An important difference in this semi-inclusive analysis compared to an
inclusive analysis is that information about the fragmentation process is
required, introducing an extra uncertainty. 

Results for the SMC final data set and preliminary results for the first
two years of HERMES running is shown in figure~\ref{fig:deltaq}. The expected 
difference of sign between up and down valence distributions is confirmed,
while the sea distributions are as yet still consistent with zero.
NLO perturbative corrections to semi-inclusive processes have been calculated
and it is possible to do a NLO global analysis of both inclusive and 
semi-inclusive data \cite{semincnlo}.

The HERMES Cerenkov detector was upgraded in 1998 to a ring imaging
Cerenkov which will provide hadron identification (i.e. ${K,p,\pi}$
separation) over the complete momentum range. While pions are 
good tags of $u$ and $d$ quarks, kaons are good tags of 
strange quarks. Thus by using kaon asymmetries
in the analysis, it may be possible to remove the constraint of the SU(3)
flavour symmetric sea and make a direct measurement of the light and
strange sea quark polarisations separately. 

An alternative method of determining the flavor decomposition of the proton 
spin, and furthermore to distinguish cleanly between valence 
($q-\bar{q}$) and sea ($q+\bar{q}$) distributions, would be to 
measure the charged current structure functions
\begin{eqnarray}
g_{5}^{W-} &=\Delta u -\Delta\bar{d} +\Delta c -\Delta\bar{s}-\Delta\bar{b},
\nonumber\\
g_{5}^{W+} &=\Delta d -\Delta\bar{u} +\Delta s -\Delta\bar{c}+\Delta{b},
\nonumber
\end{eqnarray}
at high $Q^2$ with polarized colliding beams at HERA \cite{gfive}. 
The advantages of this method is that a clean NLO analysis is 
possible; the disadvantage is that as with small $x$ measurements a 
very high luminosity would be needed to obtain useful results. 
Another possibility would be to study vector boson production at 
a hadron collider such as RHIC, with both the beams polarized \cite{RHIC}.
Again NLO calculations exist \cite{pppDY} but there are as yet no 
detailed numerical studies.

It is important to recognise that whatever method one uses, the flavor 
decomposition of first moments of quark distributions will always be ambiguous
due to the (arbitrarily large) ambiguity in the definition of $\Delta\Sigma$.
This ambiguity is only resolved if one adopts a scheme in which $\Delta\Sigma$
is a renormalization group invariant \cite{AR,SV} (i.e. an AB scheme).

If indeed the axial singlet charge $a_0$ is suppressed due to the axial 
anomaly, the effect should be target independent \cite{SV}. It might be 
possible to test this prediction by studying a variety of  
semi-inclusive processes with polarized colliding beams at HERA \cite{tarind}.
Rather dramatic enhancements are expected in some channels. 

\section{Direct Measurements of $\Delta G$}

Although it is possible to infer something about the polarization of the gluon
by the study of scaling violations in inclusive processes, it would clearly
be also useful to measure it directly by studying the final state, just as in 
unpolarized experiments. Photon gluon fusion events are 
tagged by detecting open charm production, high $p_T$ hadron 
pairs or (in)elastic $J/\psi$ production. HERMES has upgraded its 
detector so that it can better measure these channels and is 
currently taking data. The COMPASS experiment~\cite{COMPASS} has been
approved and is expected to take data from 2001/2. COMPASS has the advantage of
having a much higher energy than HERMES (200~GeV compared to 30~GeV). This
not only increases the cross section of charm production, but also reduces the
theoretical uncertainty in extracting $\Delta G$ from the measured
asymmetries. 

Perturbative calculations of cross-sections for the photoproduction of 
heavy quarks have been  at NLO \cite{photcharmnlo}, and numerical studies
\cite{photprod} suggest that the calculation is under control and that the 
asymmetries may be large enough to determine $\Delta g$ if the luminosity 
is high enough. However it must always be remembered that it is difficult 
to determine the gluon distribution from charm photoproduction even in 
the unpolarized case: the charm mass is rather low to be used as a hard 
scale, and it is difficult to disentangle resolved and unresolved 
contributions.

A similar technique is to study dijets and high $p_T$ hadrons at 
a polarized HERA. Preliminary studies at LO \cite{HERAjets} show that 
both these methods could provide a useful measurement of $\Delta g$ for 
$x\lsim 0.1$ provided only that the luminosity is sufficiently high.
The NLO corrections \cite{jetsnlo} appear to be reasonably small.

Complementary experiments in the RHIC spin physics programme~\cite{RHIC},
or scattering a polarized proton beam at HERA off a polarized fixed target,  
could determine $\Delta g$ by measuring double spin asymmetries with
prompt photon (or indeed $J/\psi$) production \cite{ppp}. At present 
calculations are only at LO, and there would presumably be
problems with intrinsic $k_T$ just as in the unpolarized measurements.
However prompt photon measurements will inevitably play an important
part in the determination of $\Delta g$ at large $x$. Dijets might also be 
useful, and here there is even a recent caculation of NLO corrections 
\cite{haddij}.
 
\section{Outlook}

The last ten years of polarized scattering experiments have seen some 
exciting and interesting discoveries, and with COMPASS, a polarized collider  
at RHIC and possibly eventually HERA, the next ten years look very 
promising too. Certainly much has, and will, be learnt about perturbative 
QCD and the distribution of polarized partons in hadrons. However 
polarizing hadron beams is an expensive business, and at some stage we 
have to ask whether these sort of experiments are likely to yield anything 
fundamentally new. Since we know already that the weak interaction is chiral,
and thus that any new physics at or around the electroweak scale is 
likely to be chiral, this may not be so fanciful as it seems. At a 
polarized RHIC it will be possible to search 
for chiral contact interations, a light leptophobic $Z'$, or right handed 
charged currents \cite{newRHIC}. Particularly interesting in the light of 
recent developments would be the search for possible chiral contact 
interactions at a polarized HERA \cite{newHERA}: while with an unpolarized 
proton beam there are two possible spin asymmetries and four charge 
asymmetries, when the proton is polarized there are a further  
twelve spin asymmetries and any number of charge asymmetries. Contact 
interactions in these channels cannot be explored with unpolarized beams.
It thus seems possible that spin might eventually play an important 
part in the understanding of new physics in the TeV region. 

\vfill\eject
\noindent
{\bf Acknowledgements:} We would like to thank G.~Court and Stefano 
Forte for critical readings of the manuscript. This work was supported in part 
by the EU Fourth Framework Programme `Training and Mobility of
Researchers', Network `Quantum Chromodynamics and the
Deep Structure of Elementary Particles', contract 
FMRX-CT98-0194 (DG 12 - MIHT). 
 

\end{document}